# Ghost interaction of breathers

Gang Xu[1], Andrey Gelash[2,3], Amin Chabchoub[4,5], Vladimir Zakharov[3,6,7], Bertrand Kibler[1]

[1]Laboratoire Interdisciplinaire Carnot de Bourgogne (ICB), UMR 6303 CNRS-Université Bourgogne Franche-Comté, 21078 Dijon, France
[2]Institute of Automation and Electrometry SB RAS, Novosibirsk 630090, Russia
[3]Skolkovo Institute of Science and Technology, Moscow, 121205, Russia
[4]Centre for Wind, Waves and Water, School of Civil Engineering, The University of Sydney, Sydney, NSW 2006, Australia
[5]Marine Studies Institute, The University of Sydney, Sydney, NSW 2006, Australia
[6]Landau Institute for Theoretical Physics RAS, Chernogolovka, 142432, Russia
[7]University of Arizona, Tucson, Arizona 857201, USA

1. Abstract

Mutual interaction of localized nonlinear waves, e.g. solitons and modulation instability patterns, is a fascinating and intensively-studied topic of nonlinear science. In this research report, we report on the observation of a novel type of breather interaction in telecommunication optical fibers, in which two identical breathers propagate with opposite group velocities. Under certain conditions, neither amplification nor annihilation occur at the collision point and most interestingly, its amplitude is almost equal to another maximum of either oscillating breather. This ghost-like breather interaction dynamics can be fully described by the *N*-breather solution of the nonlinear Schrödinger equation.

2. Introduction

The study of both formation and interaction of localized waves has been a central task in nonlinear physics during the last decades, including plasma physics, fluid dynamics, Bose-Einstein condensates and photonics. Among different types of nonlinear localized waves, solitons are the most representative and ideal testbed to investigate nonlinear wave interactions due to their intrinsic particle-like properties during the propagation [1-4]. A generic and relevant case of study for various fields of research is the elastic and nonlinear interaction of solitons described by the focusing one-dimensional nonlinear Schrödinger equation (NLSE). In this conservative and integrable system, the possible collision of solitons with different velocities does not affect their final shape or velocity after interaction, and their main physical properties keep unchanged. In general, the interaction-induced displacement in position and phase shift are independent on the relative phases of the envelope solitons. However, collision dynamics in the interaction region strongly depends on the relative phases. Consequently, in the simplest case of two-soliton collision with opposite velocities, as shown in Fig. 1(a1-d1) the two solitons appear to attract each other and cross (forming a transient peak) in the *in-phase* configuration, while they seem to repel each other and stay apart in the *out-of-phase* case. The wave magnitude at the central point of collision then evolves from the sum of the two solitons' amplitudes (i.e., amplification) to their difference (i.e., annihilation), respectively. A large range of theoretical descriptions, numerical simulations and experimental observations of such soliton interactions and their possible synchronization have been already reported [5-12].

Besides solitons, breather solutions to the NLSE are also exciting examples to investigate nonlinear wave interactions because of the salient complexities of breather synchronization related to their self-oscillating properties. From this point of view, phase-sensitive breather interactions are now widely studied [13-22]. More particularly, for co-propagative breathers, breather molecules can be formed when group velocity and temporal phase of breathers are perfectly synchronized, while for counter-propagating breathers, the phase-sensitive collision process exhibits various dynamical behaviors. Two of them have been studied in detail in the context of rogue wave formation, namely amplification and annihilation cases that look like to soliton collisions. The above interactions are fully described by *N*-breather solutions of the NLSE. However, the two-breather collision has been recently found to provide a peculiar third configuration for particular phases, neither of the above mentioned-cases, which leads to a peak amplitude at the central point of collision equivalent to the single breather amplitude before or after the collision. Phenomenologically, it seems that one breather mysteriously disappears in the nonlinear interaction region, but it then appears after that. That is why this intriguing breather interaction was vividly termed by "ghost interaction" [19]. Its generalization to the *N*-breather interaction is still under investigation. However, both detailed analysis and experimental confirmation of this remarkable dynamics for the simplest two-breather collision is still pending.

To address this, we present the observation of ghost interaction of two breathers in a single-pass telecommunication optical fiber experiment. By means of the well-known Fourier-transform pulse shaping technique applied to an optical frequency comb, we generate the initial condition for two counter-propagating breathers with desired temporal phases. The experimental results are in excellent agreement with the two-breather solution of the NLSE. We confirm that this peculiar phase-sensitive breather interaction is strictly different to the well-known soliton interactions. Our study paves the way for novel directions of investigation in the rich landscape of complex nonlinear wave dynamics.

### 3. Method

#### a. Theoretical model and breather solutions

Our theoretical framework and starting point is based on the dimensionless form of the self-focusing 1D-NLSE:

$$i\psi_\xi + \frac{1}{2}\psi_{\tau\tau} + |\psi|^2\psi = 0 \qquad (1)$$

where subscripts stand for partial differentiations. Here $\psi$ is a wave envelope which is a function of $\xi$ (a scaled propagation distance or longitudinal variable) and $\tau$ (a co-moving time, or transverse variable, moving with the wave group-velocity). This conventional form of the NLSE is widely used to describe the nonlinear dynamics of one-dimensional optical and water waves. This integrable equation can be solved by using various techniques and admits a wide class of unstable pulsating solutions known as breathers or solitons on finite background [13]. The simplest cases (i.e., first-order breathers) are well-known localized structures emerging from the modulation instability process [23]. The general one-breather solution is a localized object moving on top of the continuous wave in space-time with a particular group velocity and oscillating period. This also includes limiting cases such as time-periodic Akhmediev breathers [13], space-periodic Kuznetsov-Ma breathers [24-25] and the doubly-localized Peregrine breather [26], which have been observed in various types of experiments [27-34]. Higher-order breathers can be simply generated by considering the interaction of the above elementary breathers, thus corresponding to the nonlinear superposition of multiple breathers [13,35-36]. More generally, the NLSE has an exact *N*-breather solution, which can be constructed by appropriate integration technique by studying the auxiliary linear Zakharov-Shabat system. In the following, we restrict our work to the general two-breather solution [16,18]. It has four main parameters $R_{1,2}, \alpha_{1,2}$ (subscripts 1 and 2 correspond to the first and second breather) that control the main breather properties (localization, group velocity, and oscillation) and four additional parameters $\mu_{1,2} \in [-\infty, \infty]$ and $\theta_{1,2}$ varying between 0 et $2\pi$ that define the location and phase of each breather. More details can be found in Ref. [37]. In particular, we study the simplest one-pair breather solution $\psi_{2B}$ with breathers moving in opposite directions that can be obtained by setting $R_1 = R_2 = 1 + \varepsilon = R$, $\alpha_1 = -\alpha_2 = \alpha$. The resulting solution can be written as follows:

$$\psi_{2B}(\xi, \tau) = \left[1 + (R^2 - \frac{1}{R^2})\frac{N}{\Delta}\sin 2\alpha\right] e^{i\xi}, \qquad (2)$$

where

$$N = \left(R - \frac{1}{R}\right)\sin\alpha(|q_1|^2 q_{21}^* q_{22} + |q_2|^2 q_{11}^* q_{12}) - i\left(R + \frac{1}{R}\right)\cos\alpha[(q_1^* q_2)q_{21}^* q_{12} - (q_1 q_2^*)q_{11}^* q_{22}]$$

and

$$\Delta = \left(R + \frac{1}{R}\right)^2 \cos^2\alpha |q_{11} q_{22} - q_{12} q_{21}|^2 + \left(R + \frac{1}{R}\right)^2 |q_1|^2 |q_2|^2 \sin^2\alpha$$

In these expressions, $q_i = (q_{i1}, q_{i2})$ with $i = 1,2$ is a two-component vector function having the following components:

$$q_{11} = e^{-\varphi_1} - \frac{e^{-\varphi_1 - i\alpha}}{R}, \quad q_{12} = e^{\varphi_1} - \frac{e^{-\varphi_1 - i\alpha}}{R}, \quad q_{21} = e^{-\varphi_2} - \frac{e^{\varphi_1 + i\alpha}}{R}, \quad q_{22} = e^{\varphi_2} - \frac{e^{-\varphi_2 + i\alpha}}{R}$$

with $\varphi_1 = \eta\tau + \gamma\xi + \frac{\mu_1}{2} + i\left(k\tau + \omega\xi + \frac{\theta_1}{2}\right)$ and $\varphi_2 = \eta\tau - \gamma\xi + \frac{\mu_2}{2} - i\left(k\tau - \omega\xi + \frac{\theta_2}{2}\right)$.

The parameters $\eta, k, \gamma$ and $\omega$ are defined as:

$$\eta = -\frac{1}{2}(R - \frac{1}{R})\cos\alpha, \ k = -\frac{1}{2}(R + \frac{1}{R})\sin\alpha, \ \gamma = -\frac{1}{2}\left(R^2 + \frac{1}{R^2}\right)\sin 2\alpha \text{ and } \omega = \frac{1}{2}\left(R^2 - \frac{1}{R^2}\right)\cos 2\alpha.$$

Figure 1 (a2-d2) present the interaction of a pair of counter-propagating breathers when $R_1 = R_2 = 1.05$, $\alpha_1 = -\alpha_2 = 0.4$, thus corresponding to two identical and symmetric breathers propagating with same oscillating frequency but opposite group velocities. Here, we fixed the temporal position $\mu_{1,2} = 0$, so the central point of collision locates at the origin ($\xi = 0, \tau = 0$). According to the two-breather solution of the NLSE, we continuously vary the breather phase $\theta_{1,2}$ of over the full range $[0, 2\pi]$ to analyze its impact on the resulting waveform and amplitude at the origin. As shown in Fig. 1(b1), the amplitude of the collision-induced wave $|\psi_{2B}(0,0)|$ strongly depends on $\theta_{1,2}$ values, the maximum is obtained for $\theta_{1,2} = 0$ or $\theta_{1,2} = 2\pi$, when synchronization of the maximal amplitude of pulsating breathers is perfectly reached. When $\theta_{1,2} \sim \pi/2$, the amplitude at the central point of collision decreases to a minimum value close to the constant background amplitude $|\psi_0| \sim 1$. Interestingly, there is another local peak of $|\psi_{2B}(0,0)|$ at $\theta_{1,2} = \pi$, whose amplitude is very close to that of a single breather before or after the collision $|\psi_1| \sim 2.7$. In order to better reveal the space-time dynamics of such breather interactions, we present the full wave evolution in Fig. 1(b2-d2) for the following cases: (i) $\theta_1 = \theta_2 = 0$, the synchronized collision of breathers that generates a rogue peak with extremely high amplitude (already reported experimentally in Ref. [15]); (ii) $\theta_1 = \theta_2 = \pi/2$, the quasi-annihilation of breathers that gives rise to very small perturbation located on the continuous wave (already reported experimentally in Ref. [19]). However, note that in this case, we observe a spatial symmetry breaking phenomenon before and after the collision of these two breathers because of the noticeable $\pi$-phase shift (see Fig.1(c2)). This specific configuration of breather collision termed as superregular breathers can be regarded as a prototype of small localized perturbations of the plane wave for describing modulation instability [16]; (iii) $\theta_1 = \theta_2 = \pi$, the two breathers are almost transformed into a single one in the main interaction region, at the origin $|\psi_{2B}(0,0)|\sim|\psi_1|$, which raises the impression that one breather has vanished (see Fig. 1(d2)).

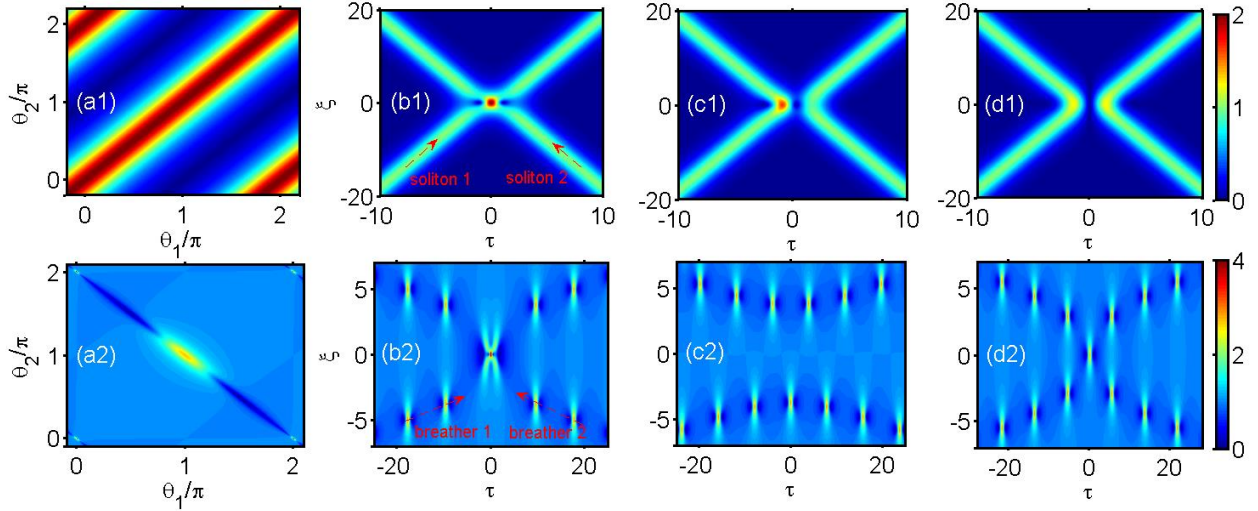

**Figure 1: Typical temporal evolution of soliton-pair interaction (first line) and breather-pair interaction (second line).** (a1) Dependence of amplitude at the soliton collision point $|\psi_{2S}(0,0)|$ on the soliton phases $\theta_1$ and $\theta_2$. (b1-d1) Amplitude evolution of soliton collision with soliton phases: $\theta_1 = 0, \theta_2 = 0$ (b1); $\theta_1 = \pi/2, \theta_2 = 0$ (c1); $\theta_1 = \pi, \theta_2 = 0$ (d1). (b1-d1) are plotted based on the two-soliton solution of NLSE with the soliton parameters: angular frequencies $\Omega_1 = -\Omega_2 = 0.5$; soliton amplitudes $A_1 = A_2 = 1$. (a2) Dependence of amplitude at the breather collision point $|\psi_{2B}(0,0)|$ on temporal phases $\theta_1$ and $\theta_2$. Prototypes of interactions include Amplification (b2), Annihilation (c2), and "Ghost interaction" (d2). (a1-d2) are plotted based on the one-pair breather solution of NLSE. In all these cases, key parameters of breathers are listed as follows $R_1 = R_2 = 1.05, \alpha_1 = -\alpha_2 = 0.4, \mu_1 = \mu_2 = 0$. while $\theta_1 = \theta_2 = 0$ for (b2); $\theta_1 = \theta = \pi/2$ for (c2) and $\theta_1 = \theta_2 = \pi$ for (d2). Red arrows in (a2) and (b2) indicate the moving motions of solitons and breathers respectively.

We would like to emphasize that the ghost interaction of breathers as illustrated in Fig. 1(d2) cannot occur for the soliton counterpart (see Fig. 1(a1-d1)). To clarify this point, we compare systematically the phase-dependent soliton collision and the phase-dependent breather collision. Similarly, we consider a pair of counter-propagating solitons with the amplitudes $A_1 = A_2 = A$ and the frequencies $\Omega_1 = -\Omega_2 = \Omega$. In this situation, the two-soliton solution on zero background can be written in the following form [18]:

$$\psi_{2S}(\xi,\tau) = 2A \frac{\Omega^2(|q_1|^2 q_{21}^* q_{22} + |q_2|^2 q_{11}^* q_{12}) - iA\Omega[(q_1^* q_2) q_{21}^* q_{12} - (q_2^* q_1) q_{11}^* q_{22}]}{A^2 |q_{11} q_{22} - q_{12} q_{21}|^2 + \Omega^2 |q_1|^2 |q_2|^2} e^{i\xi}. \qquad (3)$$

In this expression, $q_i = (q_{i1}, q_{i2})$ with $i = 1,2$ is a two-component vector function having the following components: $q_{11} = e^{-\varphi_1}, q_{12} = e^{\varphi_1}, q_{21} = e^{-\varphi_2}, q_{22} = e^{\varphi_2}$, with

$$\varphi_1 = \frac{A}{2}\tau + \frac{A\Omega}{2}\xi + \frac{\mu_1}{2} + i\left(\frac{\Omega}{2}\tau + \frac{\Omega^2 - A^2}{4}\xi + \frac{\theta_1}{2}\right) \text{ and } \varphi_2 = \frac{A}{2}\tau - \frac{A\Omega}{2}\xi + \frac{\mu_2}{2} + i\left(-\frac{\Omega}{2}\tau + \frac{\Omega^2 - A^2}{4}\xi + \frac{\theta_2}{2}\right).$$

Again, the $\mu$ and $\theta$ are responsible for soliton position and phase. We let $\mu_1 = \mu_2 = 0$ and Fig. 1(a2) demonstrates the dependence of the amplitude at the collision point $|\psi_{2S}(0,0)|$ on $\theta_1$ and $\theta_2$.

Compared to the breather collisions, here the key parameter for soliton collision is the relative soliton phase $\theta_1 - \theta_2$. In general, amplification interaction occurs for $\theta_1 - \theta_2 = 0$, and annihilation interaction happens for $\theta_1 - \theta_2 = \pi$. While for other values of relative soliton phase, $|\psi_{2S}(0,0)|$ keeps being low (~0), and a partial energy exchange takes place from one soliton to another in the collision area which leads to a remarkable time-parity symmetry-breaking (examples shown in Fig. b1-d1).

### b. Experimental setup

In order to validate these theoretical predictions about ghost interaction of breathers, we perform experiments with light waves by means of high-speed telecommunication-grade components as depicted in Fig. 2. The main challenge here is the arbitrary wave shaping to obtain the specific initial excitation of counter-propagating breathers with optimized relative phases. To this end, a 20-GHz optical frequency comb (more details can be found in Ref. [20]) passes through a programmable optical filter (wave-shaper) to precisely control both amplitude and phase characteristics of each comb line. As a result, we can synthesize any arbitrary perturbation of a continuous wave background in a time-periodic pattern whose frequency is equal to the comb spacing. This temporal pattern is then amplified by erbium-doped fiber amplifier (EDFA) to achieve the exact excitation of the two-breather solution in terms of average power for nonlinear propagation into our single-mode optical fiber (SMF).

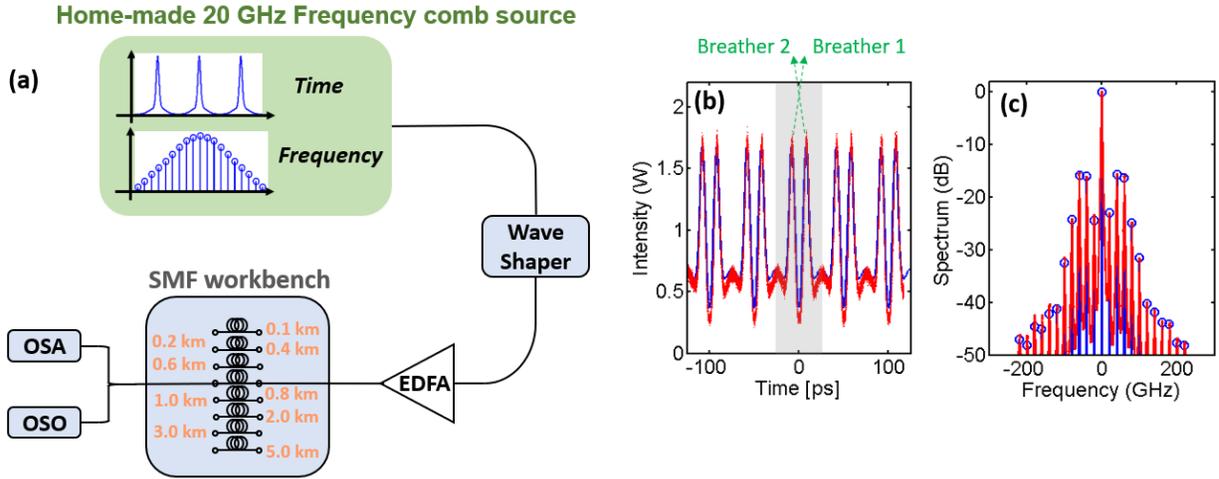

**Figure 2: Experimental setup and generation of initial conditions.** (a) Schematic diagram of the experimental setup. EDFA: erbium-doped fiber amplifier; SMF: single mode fiber; OSA: optical spectral analyser; OSO: optical sampling oscilloscope. Shaded-green box represents the home-made frequency comb source with a repetition rate of 20 GHz. (b-c) Designed initial conditions at 20 GHz repetition rate for a pair of counter-propagating breathers in both the temporal and the spectral domains. Solid blue lines are theoretical curves; Solid red lines are experimental measurements. Here breather parameters are: $R_{1,2} = 1.5$, $\alpha_1 = -\alpha_2 = 0.5$, $\mu_{1,2} = 0$, $\theta_{1,2} = \pi$.

The corresponding temporal and spectral power profiles of the light-wave are presented in Fig. 2(b-c). Note that the initial condition for the breather pair is time-periodic with a period of 50 ps. Hereafter, we select the center time slot ($-25$ ps $< t < 25$ ps) to investigate the collision dynamics of the breather pair as shown in grey shaded area in Fig. (2b). The nonlinear propagation is studied with different lengths of the same fiber and characterized by means of an optical sampling oscilloscope (OSO) with sub-picosecond resolution in the time domain and a high dynamics-range optical spectrum analyzer (OSA) in the Fourier domain. The maximum propagation distance fixed was chosen to limit the impact of linear propagation losses in our optical fiber as well as possible interaction occurring between neighboring elements of the periodic pattern. Our fiber properties are the following: group velocity dispersion $\beta_2 = -21.1 \text{ ps}^2\text{km}^{-1}$, linear losses $\alpha = 0.2 \text{ dB km}^{-1}$, and nonlinear coefficient $\gamma = 1.2 \text{ W}^{-1}\text{km}^{-1}$.

## 4. Results

We present our experimental results on the nonlinear space-time evolution for the breather pair studied in the above theoretical section, for the specific temporal phases $\theta_1 = \theta_2 = \pi$. To this purpose, we fixed the average power to $P_0 = 0.74 \text{ W}$. Then we gradually increase the propagation distance (i.e., the fiber length) by a step of 100 m. One can retrieve the correspondence between normalized and physical units by making use of the following relations between dimensional distance $z$ $(m)$ and time $t$ $(s)$ with the previously mentioned normalized units: $z = \xi L_{NL}$ and $t = \tau t_0$. In these expressions, the characteristic (nonlinear) length and time scales are $L_{NL} = (\gamma P_0)^{-1} \sim 1216 \, m$ and $t_0 = \sqrt{|\beta_2| L_{NL}} \sim 4.74 \, ps$ respectively. The dimensional optical field $A(z,t)(W^{1/2})$ is $A = \sqrt{P_0}\,\psi$.

Figure 3 (a1-a2) presents the concatenation of temporal (amplitude) profiles and power spectra which were recorded at the output of the distinct fiber segments with increasing length. The careful control of phases allows to observe the ghost interaction between the counter-propagating breathers. The full space-time dynamics is indeed in excellent agreement with theory shown in Fig. 3(b1-b2). One can notice the five mains peaks appearing during the whole evolution studied in Fig. 3(a1): two peaks at $\xi_1 \sim -2.2$ for the two breathers before collision; one peak at $\xi_2 \sim 0$ at the collision point; and two peaks at $\xi_3 \sim 2.2$ for the two breathers after the collision. Correspondingly, we observe the maxima of spectral broadening for respectively $\xi = \xi_1$, $\xi = \xi_2$ and $\xi = \xi_3$ (shown in Fig. 3(a2)), thus confirming the different nonlinear temporal focusing patterns. Figure 3(c1) presents the comparison of the recorded temporal waveforms for $|\psi(\xi=\xi_1,\tau)|$, $|\psi(\xi=\xi_2,\tau)|$ and $|\psi(\xi=\xi_3,\tau)|$. Strikingly, all these five peaks are found to nearly exhibit similar waveforms and maximum amplitudes, this is also corroborated by the spectral analysis reported in Fig. 3(c2). Only slight discrepancies can be noticed mainly ascribed to the linear propagation losses in our optical fiber and some artefacts of the initial wave shaping.

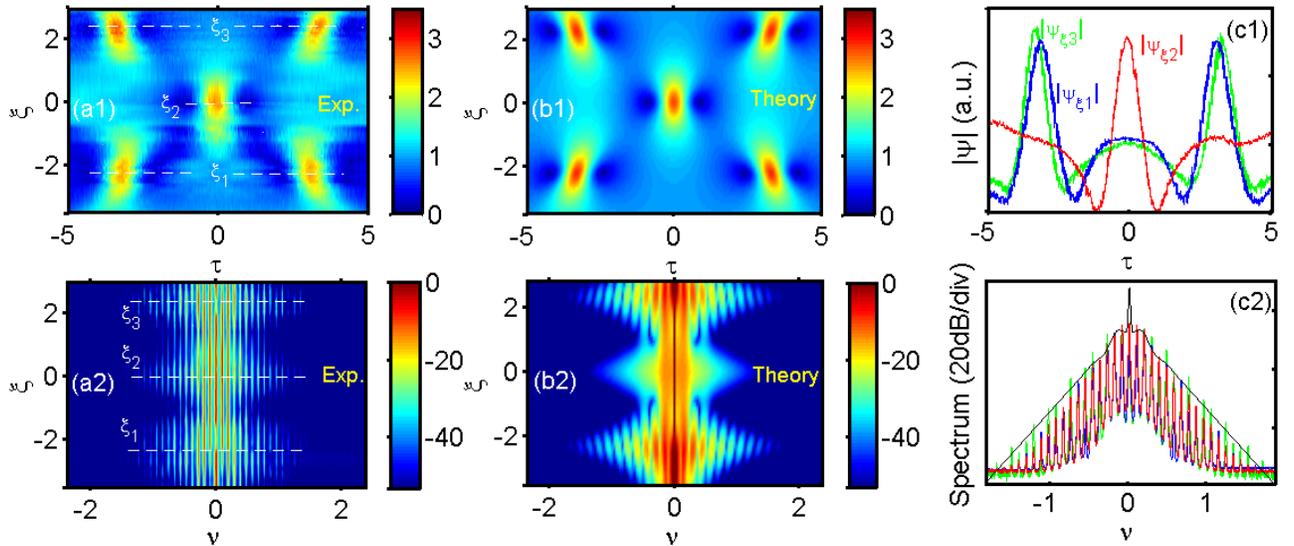

**Figure 3: Experimental observation of ghost interaction of two breathers.** Colour map showing the evolution of both temporal (a1) and spectral (a2) profiles for the two breathers observed in experiment. Dashed white lines indicate the position of local maximum amplitude, which are also the position of largest spectral broadenings, before collision ($\xi_1 \sim -2.2$), during collision ($\xi_2 \sim 0$) and after collision ($\xi_3 \sim 2.2$). (b1-b2) Corresponding theoretical predictions based on the two-

breather solution of NLSE. (c1) Comparison of the amplitude profiles measured at $\xi = \xi_1$ (blue curve), $\xi = \xi_2$ (red curve) and $\xi = \xi_3$ (green curve). (d) Comparison of power spectra recorded at $\xi = \xi_1$ (blue curve), $\xi = \xi_2$ (red curve) and $\xi = \xi_3$ (green curve). Thin dark curve is the theoretical spectrum at $\xi = 0$. Key parameters of the breather pair: $R_{1,2} = 1.5$, $\alpha_1 = -\alpha_2 = 0.5$, $\mu_{1,2} = 0$, $\theta_{1,2} = \pi$.

## 5. Discussions

As shown above, during the ghost interaction of the two breathers, only a single breather peak remains at the collision point. The reason for this intriguing phenomenon is that there is a continuous power exchange between the background and each localized perturbation all along the propagation, which is an intrinsic property of breathers. Therefore, when these two breathers nonlinearly interact near the collision point, for particularly initial phases, one of the breather peaks is almost hidden in the background and then appears again after the collision without raising any issue of energy conservation. Moreover, the breather pair keep the spatial and temporal symmetry for the whole evolution. It is worth to mention that such a peculiar ghost interaction does not occur in conventional soliton-soliton collision scenarios because of the lack of pulse-background exchange [see Fig. 1(a1-d1)].

In summary, we presented different configurations of phase-sensitive breather collisions, especially, the very intriguing type of ghost interaction. Our experimental observations fully confirmed the theoretical predictions based on the exact two-breather solution of the NLSE. Our study was restricted to the interaction of two identic counter-propagating breathers, while much more complicated many-body interactions of breathers with asymmetric conditions, including different amplitudes and/or oscillating frequencies, still require further and more complex investigations. Our results represent a novel step towards the understanding of interactions between localized waves in nonlinear physics. It may naturally lead to further relevant experimental studies and theoretical investigations in various fields of wave physics.

## 6. Funding

French National Research Agency (PIA2/ ISITE-BFC, Grant No. ANR-15-IDEX-03, "Breathing Light" project). Theoretical part of the work was supported by Russian Science Foundation (Grant No. 19-72-30028).